# Fixed interfaces, adaptive interfaces... What is next?
# Total movability – a new paradigm for the user interface.[*]

*Abstract.* Users can't talk with computers in their natural language (machine codes), so there are interfaces that allow such communication. 40 years ago the outcome of computer programs was in the form of long listings covered by numbers and even the format of those numbers was determined by developers. Throughout the latest 25 years: program views and results are shown in a wide variety of shapes and variants, but all these possibilities are predefined and fixed in code by developers; nothing outside of their approved solutions is allowed. My vision from now on into the future: developers are responsible only for correct work of a program (calculations, link with the database, etc.) and suggest a good default interface, but not determine all possible scenarios; only users decide WHAT, WHEN, and HOW to show. This will be a revolution in our dealing with computers, but there are obvious questions. How this step can be made? Do all users need such change? Is it going to be a burden for users or a welcome revolution?

We can't talk with computers in their natural language, so there are interfaces that allow our communication. At the beginning there were big computers and very few people who knew how to talk to them. When the term *personal computer* didn't even exist, the overwhelming majority of programs were scientific or engineering. The researchers who worked on those projects were mostly proficient in math; they easily learnt FORTRAN and turned their knowledge into algorithms and codes. At those days it was common for the same person to be a researcher, an author of the code, and the analyser of results. In case of need, the same person could change the format of the output numbers.

In a while you could see more specialization in programming area with different people being researchers and program developers, but at that time researchers and programmers often worked in close collaboration. If anyone without sufficient knowledge of programming needed the results of long and tiresome calculations, then this person had to explain his problem to those sorcerers who would produce some magic and give back the needed results. Programmers communicated with computers in one of secret languages that only members of their guild knew. The collaboration between researchers and programmers was so close that any requirements on changing the output were solved after a short discussion on a personal level. The interface was occasionally changed by programmer, but at any moment it was fixed and absolutely controlled by developer.

Things began to change rapidly with the beginning of PC era and especially for programs used by thousands or millions. It's impossible to imagine that all those users would be absolutely satisfied with the designer's ideas of interface and that became the huge problem. The solution to this problem was obvious: give users an instrument to adapt the interface to their demands. For the last 30 years the design of interfaces for computer programs was influenced mostly by the ideas of *adaptive interface*. Those numerous ideas are described in hundreds (more likely – thousands) of papers and books and are seen in all currently used programs. Now there is simply nothing except adaptive interface; it became an axiom, a dogma. With all its excellent results, it contains a fundamental flaw hidden deep inside and never publicly discussed: adaptive interface is based on the postulate that designers know better than any user what is really good for each and all situations. The adaptive interface is usually called friendly because it gives users a choice; however, any selection can be made only among the possibilities which beforehand were considered as appropriate by developers and only these variants are allowed. The interface is not fixed any more; it gives users some choices but it is still controlled by developer.

When users are not satisfied with a set of allowed solutions and demand something different, then a new set of possibilities is coded or another instrument to select among them is developed. The number of commands and possibilities in new programs increases; the interface becomes so sophisticated that… This situation is perfectly described in the preface to [1]: "*You have to figure out how to cast what you want to do into the capabilities that the software provides. You have to translate what you want to do into a sequence of steps that the software already knows how to perform, if indeed that is at all possible. Then, you have to perform these steps, one by one.*" As a result, users often are not doing exactly what they – users –want to do; instead, they try to find how to ask an application to do something as close to their need as possible.

For many years I design very complicated programs; mostly for different areas of science and engineering (speech analysis, telecommunication, analysis of electricity networks, thermodynamics) but not only. All those programs were and are used by specialists in each particular area; those people try to solve some very complicated tasks, so they work with a lot of parameters to regulate the analysed processes. The programs include the tuning of all those parameters, so the interfaces of those programs have to be very sophisticated. Several years ago I understood that there was something common in the design of sophisticated programs for many different areas: there was an obvious stagnation in design and this stagnation

---

started approximately 15 years ago.  I tried to look into the cause of this stagnation and found out that it was in the dominant design philosophy – in the basic idea of adaptive interface.

Research work is usually done with some specialized programs.  The applications are developed at the good level and are supposed to be used in different cases; such programs try to predict and implement the most needed scenarios.  The designers of those programs can be good or not so good (I am familiar with many strange cases), but they are always not as good specialists as the researchers who have to work with those programs.  Researchers are the best specialists in each particular area; developers are either not so good or significantly worse.  And we have the biggest paradox for all areas of the scientific and engineering applications: the best specialists have to work under the restrictions imposed on them by significantly less specialists.  It happens so because there is no other idea for design of complex programs than the popular adaptive interface; under this dogma all the possible variants are approved and coded by the developers and no one can get anything else from those programs.

For me the situation with the dominant role of adaptive interface in design of interfaces and the fury (this is the most correct word) with which some specialists defend the divinity of adaptive interface looks like the situation with the guilds in Middle Ages.  Only the members of the guild could put on the market any specialized product and each of those products was strictly standardized; no changes or improvements were allowed.

The trend in producing any goods in any particular area is always the same throughout the ages: first very primitive products with the predetermined features, then the production of several variants to satisfy different groups of customers, and then the widest line of products through which customers can search for their personal satisfaction.  Looks like in interface design we have now even the reverse movement, because in our days the most promoted form of adaptive interface – dynamic layout – gives users not even the selection among several choices but only a single solution! And users of such programs have no other choice but work under this imposed variant.  Certainly there are prophets who declare that dynamic layout is the best form for interface design and explain its benefits.  Not long ago I got such an explanation in a private letter from one of the MIT professors: "*Without dynamic layout, the end user would have to manually, one by one, resize and reposition the elements inside.  So dynamic layout does confer usability benefits by making the user interface more efficient: one resize action by the user results in many automatic resizes and repositions of dependent objects.*"  In the same way somebody can declare that slavery is the best form of social organization because slaves don't need to think about food or shelter; slaves are provided with both.  Do I need to remind about some negative sides of slavery?

Users are told all the time that designers know better than anyone else how the program must look in all the situations.  If you want to solve your task on the proposed application you have in exchange to give up the control over the interface to designers and absolutely rely on their decisions.  Is it going to be so forever?  What about another division of responsibilities: developer provides the correct work of an application (calculations, saving and restoring the parameters, printing, and etc.) while user, if he wants, can easily change the interface.  Such model can work under several conditions:

- User is going to change the interface only on his own wish, so a good default view is provided by developer.

- Each user can change the interface in an arbitrary way, so there are no predefined variants.  Without such variants, a developer is taken out from the interface change; he is simply banned from this process.

- Those changes of interface must be simple and natural; no instructions are needed.

What feature must be implemented to make such a revolutionary step?  To understand the new idea, let's consider the analogue situation.  Suppose that you have rented for some time a house with the furniture, fully equipped kitchen, and all other needed things.  The owner tried to do the best and before your arrival put everything in such an order that, from his point of view, would be the best for you.  On your arrival, you can be satisfied with all you see and keep everything in the same order, but chances are high that you will move some pieces of furniture and other things around the house according to your own preferences.  And throughout your stay at the house you will continue to move the things around whenever you feel any need for it.  Though it is your temporary house, you want to feel comfortable at any moment and this can be easily achieved by moving things around.  Those movements can be caused by many different things: rainy or sunny day outside and your desire to have more or less light, a home party would require more free space while a private conversation would need two armchairs close to each other, a dog would need some special corner, and so on.  When you leave the house, another tenant will arrive and will move the things around the house according to his preferences.  The same circumstances would result in similar actions by different tenants; similar but not identical, because everyone has his own taste.  The solution to a complicated enough problem of organizing a comfortable living for every tenant is easy enough and doesn't require a special course: everyone knows that all things are movable and can do it himself.  As much as he wants and whenever he feels any need for it.

In this case of renting we have the main task (house for rent) and many users (tenants are coming and going) with their preferences.  There is also an easy way to make everyone comfortable by moving the things around at any wish. Do you see the analogy with the programs?

My vision of the future interface design is based on taking the full control over the interface from developer and giving it to users; to achieve this I implement the movability of all the screen objects. This can significantly affect users' work with applications and will change the developers' work. Is it so important for all users or not? Are all the programs have to be changed or not? How significant must be the changes in development? I'll return to these questions further on, but let's start with a glimpse on the proposed idea.

In all our programs, there are buttons and menus to transform our commands, lists for selection of options, special boxes to read or write texts, areas to show pictures, and so on; each screen element is used for special operation. Actions, associated with those elements, are the same regardless of the sizes and positions of the screen elements, but in currently used programs these parameters are decided by designers. If the screen elements would be movable / resizable by users throughout the time when an application is running, then such moving / resizing would not affect the work of application at all but would allow each user to have the interface he personally wants. If any element and group of elements can be easily moved, resized, and reorganized by any user at any moment and in the simplest way, for example, with a mouse, then a program can be personalized by each user in seconds without any loss in its work.

There are millions of users; their demands are so different that there is no way to satisfy all of them, but here are three main groups into which all users can be classified.

- I don't bother about interface problems and I don't want to make even a single change in interface myself; I can adjust to anything if the program solves my problems.

- I can spend some time on adjusting the program to my problems, but I want this time to be minimized.

- I am going to spend a lot of time working with this program, so there must be a very flexible instrument of adjusting it to my changing problems; I am ready to use this instrument, but it must be not overcomplicated.

For the first group the new applications work exactly as the existing so these users are not bothered at all. For others the benefits are correlated with their efforts. You can significantly improve the view in one or two seconds or you can organize such needed view which no developer would think out beforehand or allow. Even such changes would take only seconds, but you get what you really want and need.

Adaptive interface is aimed at satisfying an average user; in this way it is the worst scenario for the most complicated applications (scientific, engineering, and similar) and for the most experienced users who work on solving the most crucial new tasks; instead of solving the problems such users spend a lot of time fighting with or trying to deceive an application they are working with.

For the average customer the use of adaptive interface is also not an excellent solution: aimed at an average (hypothetical) user, in reality it satisfies nearly nobody. It's exactly like a well known mathematical problem of producing the best function for a big number of experimental data: it is determined by minimizing the sum of square deviations, but the graph rarely goes exactly through any of the data points.

Not average user works with an application but thousands or millions of real people. Each one of them has his personal view on how a program must work and look. Is there any way to satisfy each of those users? My answer is that the movability of all the screen objects is a huge step in solving such a problem. But before going further on, I want to underline two main things about the discussed movability.

First, I am not writing about movies or games in which the movements of elements are organized according to the previously developed scenario. I am writing about the movements which can't be predicted and are totally initialized and controlled by users. Developers only provide the elements which can be moved by users, but everything is controlled by users only. This means that users decide WHAT, WHEN, and HOW must be shown on the screen. Thus users get the full control over applications; programs become *user-driven*.

Second remark is about the novelty of the proposed idea. Whenever I mention that an idea of total movability of screen objects is absolutely new, I immediately get a response that this is wrong because, for example, everything is movable in Windows. Yes and no, because there are two different levels. At the level of multi windows operating system users have the full control over the elements, but there are only two types of elements. Users can move and resize windows; they can open the new windows, close the unneeded, and place the windows associated with the programs in whatever way they want. Users can also move the icons around the desktop. There are no other elements at that level; thus users have total control at this level. The situation inside the applications is absolutely different because there we have an infinitive world of different objects born by the imagination of millions of developers around the world. Start any application and try to move any object inside. With an exception of few very special programs (like *Paint* and similar) nothing can be moved inside the applications. At the inner level we have not only rectangular objects but elements of the arbitrary shape; to organize users' control over all the inner elements you need some algorithm of turning any element into movable. Such algorithm was never invented before and was never demonstrated.

Were there previous attempts to provide the movability of objects by users inside the working programs? Movability of objects in our everyday life is so natural that it would be natural to expect the works on movability of screen objects in the past. There were such examples and all of them can be divided into two separate groups.

The first one includes the systems in which, instead of real movability, you have a set of allowed positions, i.e. a classical adaptive interface. Some specialists like to say that movability was implemented years ago, for example, in the Morphic system. This statement is absolutely wrong. A good description of Morphic says that that iconic system gives only three possibilities for lining the elements ([2], page 22). It was also implemented in exactly the same way before Morphic (see references in [2]).

The second group of solutions goes back as far as SketchPad by I.E.Sutherland [3]. The main idea is to have somewhere in memory a description of a drawn contour and change the coordinates of a figure according to the movement of some device. Few readers were around to look into those codes 50 years ago, but nearly everyone is familiar with the movement of dotted rectangle in *Paint*; the idea is the same. You have the stored coordinates of a contour; when you press a mouse somewhere in the vicinity of a line, the mouse coordinates are compared with the stored coordinates of lines and the "pressed" one is determined. When the mouse is moved without button release, the change in figure's coordinates is calculated according to this movement. Movement of a figure is demonstrated by erasing the old lines and painting the new; twenty and plus years ago it was perfectly accomplished by using XOR operation.

From time to time you need the movable objects in your application and occasionally good programmers organize such movability. For each class of movable objects the special algorithm is used; such algorithms are complicated and never used for anything else. I have done it 15 and 20 years ago in some of my programs and those old algorithms could be used only for those specific objects. We need the movability of screen objects because it opens to us absolutely new world of program design, but the problem exists: how to turn into movable any screen object of an arbitrary shape and complexity?

The main problem was in development of an algorithm that could answer several demands:

- Applicable to any object.
- Easy in use for developers.
- Easy and natural for users.

Several years ago I thought out an algorithm that is based on using the invisible cover for each screen object. Those objects can be of an arbitrary shape, origin (graphical objects and controls), and complexity; **figure 1** demonstrates only some of the popular shapes. Cover of each object consists of a set of elements which I call nodes. Because of the huge variety of shapes of screen objects, there is a question of how many different types of nodes are needed. I found out that it was enough to have three types of nodes – convex polygons, circles, and strips with rounded ends; the size of any node is not limited.

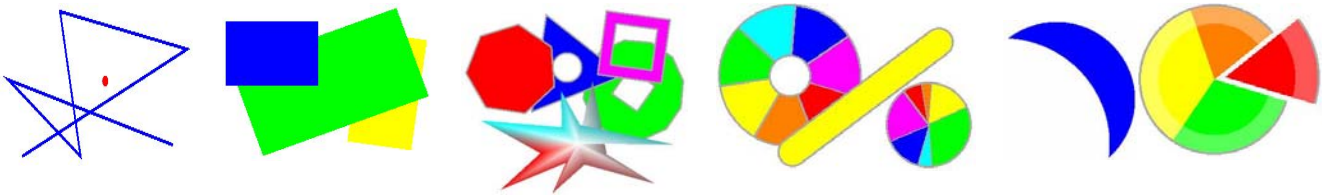

**Fig.1** Some most popular shapes of objects

Each node is used either for movement of the whole object or for its resizing (reconfiguring). The overlapping of those nodes is not a problem at all, so there is no need to cover an arbitrary area with those nodes in a very strict way. On the contrary, the use of nodes overlapping allowed to develop an easy and elegant technique for moving / resizing of objects with the curved borders and with holes. Algorithm is easy to use in programs; for better understanding there is a book [4] accompanied by a huge Demo application with more than 100 examples and all its codes available. It's like the mathematical analysis: you have to understand the ideas of differentiation and integrals; you have to know several often used techniques, and on this basis you can start to solve any of your problems. Everything else depends on your brains, your skills, and the spent time. Simple elements from **figure 1** are used in the book to demonstrate the code for forward movement, rotation, for synchronous and related movements of objects. There are also a lot of examples of real applications from different areas. Though I started to work on movability for scientific applications, it turned out to be of highest need in the programs from many different areas.

From users' point of view, the easiness of moving / resizing of objects is based on three simple rules.

- Everything is movable.
- An object is moved by any inner point.

- An object is resized by any border point.

If these rules are used in all the applications and for all the objects without exceptions then there is no need in any instructions.  I'll demonstrate with the next two examples what the application of these algorithm and rules can do in normal complicated programs.

The first example works with personal data.  From time to time nearly everyone has to deal with some application which asks us to type in some personal data, stores this data somewhere in database, and can display the stored data later on request.  If you have to deal with such an application once in several years then you have a chance to live through each of those encounters regardless of the implemented interface, but there are people, like HR personal, who have to work with such programs for many hours every day and for those people the design of such application and the easiness of its change according to personal taste and current task at any moment are crucial.

Personal information can include a wide variety of data.  Let us decide that in our case the maximum set of data about any person includes a name, date of birth, address, contact information, and some information about professional activity; **figure 2** shows the default view of this application.  When such program is linked with a data base and you type in the name of a person, then all the informative controls in view are filled with the data about that person.  This application can be used for many different tasks; in each of them the really needed part of information is different.  The screen space is always very valuable, so it would be a high quality program if any user can rearrange it in seconds in such a way that at any moment it will show exactly the required part of information and nothing else.  Whatever is needed and what is considered unnecessary is decided only by user.  The number of possible variants is infinitive, so there are no chances that any form of adaptive interface with its predefined scenarios would answer such a task.  For user-driven application it's an easy task and anyone can set the needed view.  **Figure 3a** shows the same application for the period when Christmas cards have to be sent, while **figure 3b** demonstrates the view which is more suitable for organizing some professional meeting.  Switch to any of these views takes only a couple of seconds; the default view can be reinstalled in an instant via a command of context menu.

**Fig.2**  The default view of *PersonalData*

**Fig. 3a**                                           **Fig.3b**

You remember that example of the rented house and tenants with different requests?  Now look at the default view of our *PersonalData* application (**figure 2**) and two of its customized views (**figure 3**).  Application is an analogue of a house, but the flexibility of its rearranging is certainly much higher than for real objects around us.  We can move a refrigerator from place to place, kids can even repaint it, but it would be a problem to change its size, though a lot of people would like to have such a model.  The `TextBox` for the street address in the last application is our "refrigerator" and it will continue to show the address regardless of the place of a control, its size, the used font, or the color.  If at some moment this `TextBox` is not needed then it can be taken out of the screen only to be reinstalled in an instant on the first wish.  The algorithm of movability allows to use the parallel worlds as an unlimited store for temporary unneeded objects.  All the changes can be done in seconds; different changes can be applied to a single element, or a group or elements, or a whole view of the program, but each control will continue to show the exact type of information with which it is associated.  Developer guarantees the correctness of links between data base and all those controls, while users have an unlimited freedom of organizing the view in any way they want.

When such application is designed under the ideas of adaptive interface, then developers put huge efforts in organizing several typical views which will be good enough for the majority of users.  In reality each user will select that of the allowed configurations which is as close as possible to his task at the particular moment.  With the movability / resizability of all the screen elements each user gets not the good enough but exactly the needed configuration.  The efforts on rearranging the view in both cases are identical, but instead of the suitable enough view you get exactly the view you want.

Another program is the *Analyser of functions*; its only limitation is on the type of functions with which this analyser works: they can be either usual Y(x) functions or parametric functions, described by the pair {X(r), Y(r)} (**figure 4**). This *Analyser* can be used either for education or for quick analysis of the needed functions, but the program itself is very close in ideas and implementation to other applications which I have developed for real work in the Department of Mathematical Modelling. Not long ago I wrote in [5] about the way in which such user-driven applications changed the scientists' work.

For a long period of time scientists worked only with pen and sheets of paper; the invention of big computers lessened the burden of long and tiresome calculations and added to the things on our desks the numerous listings with the results of those calculations. Yet, the paper work was still the main thing and for this process anyone had his own habits.

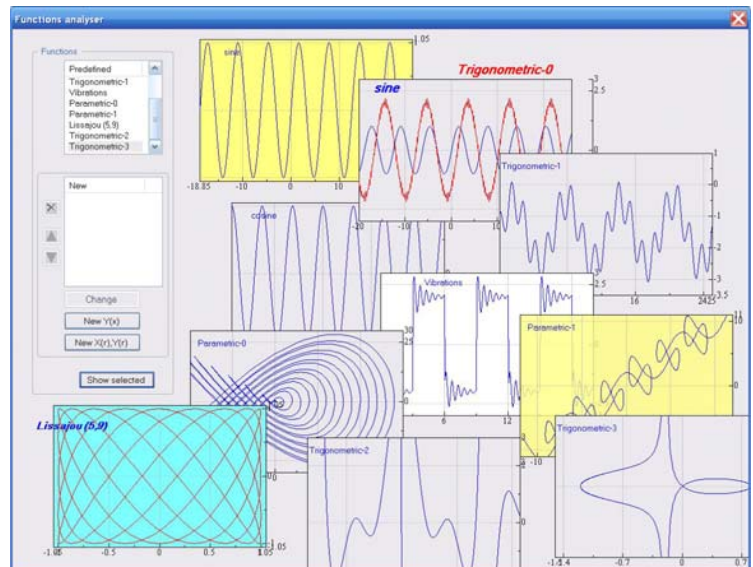

**Fig. 4** An application to demonstrate functions

Some people accurately pile the sheets with their paper work; others prefer to throw them around. Each one has his own habits and these habits serve in the best way for their owners. These habits are very personal and provide the most effective way of work for each one.

Personal computers changed the way of work for millions of people. At the same time each program puts the same restrictions on all its users by allowing everyone to work only inside the boundaries of the developers' vision. User-driven applications return the freedom of personal choice to the type of work which, by its origin, is very individual and demands the individual organization. Even this would be a huge result of implementing the movability of the screen objects, but this is not all. Not a coincidence but another great result of implementing movability for all and each object that it allows to design the scientific / engineering applications which simply can't work without it. Throughout the last two or three years I have developed for my colleagues several applications about which they were thinking for years but didn't see any even relatively good way for their implementation. With the movability of all elements and parts it turned out to be possible on the highest level; some of these programs are demonstrated and discussed in [4].

It is really interesting to watch the researchers' work with the applications on movable elements. New plots are organized in order to look at more and more results; the most interesting results are moved on top and enlarged, while less important can be moved aside, squeezed, or deleted. It's exactly like an old work with sheets of paper with the graphs; researcher goes through them trying to analyse and undercover results of experiments. On those old days there were drawing, redrawing, visual comparison, throwing away the wrong graphs and so on. Now researchers can do all the same things with the screen objects plus other actions provided by computer: zoom, printing copies, changing visibility parameters for better comparison, saving and later restoring. The most important thing in such a work is the fact that the research work and analysis are no more limited by whatever one or another developer of a program considered to be more than enough for researchers.

By developing such applications, I put no pressure on my colleagues in the way they have to do their research work. I know how to develop applications for them to analyse the results of experiments and do not put my nose (or hands) into the problems of hydrodynamics in which they are the first class specialists. You can look at such revolution in design of programs (and not only engineering or scientific programs) from such point of view. What is the best possible scenario in people's dialogue with computer? For example, you need to solve some problem and to do this you start giving some initial commands, set the input data, and define some parameters and limitations. This is called the *input data* stage. The real *calculations* are done by the previously coded methods provided by developers, but there are a lot of tuning parameters for these methods and the selection of these parameters depends on the users' qualification. Then you get the results (*output* stage) and… the most interesting thing starts. In normal (complex) task you never receive the result as one number but rather as a big or very big collection of data that you have to analyse and make a conclusion. This, to very high extent, depends on the qualification of user, but in any case the set of results must be demonstrated to user in the best personal view. No designer can decide what view is the best for each particular user and can provide only a good view for an "average" user. With the implemented movability, and if it is organized in a very simple way, then each user can set the preferable way of demonstration for himself. That is the main idea of the best applications: to cut the developers' control over the visualization and to pass this control to users. Not inside the limitations enforced by developer but absolute control; at the same time this control over visualization must be not a burden for researchers, but this control must be organized in an absolutely natural way.

## Changes in development and use

The user-driven applications are undistinguishable from the currently used but they differ in development and use. For developers, there are several rules of design.

1. All the elements are movable.

   There are no exceptions; all the objects, regardless of their shape, size, or complexity must be turned into movable. If you decide to add movability to nearly everything **but** this and that, then uses will bump into these hillocks on the road again and again. With an adequate thought about you as a designer.

   Movability of elements has a very interesting feature: it's enough to make movable some element of the form and then the logic of application will require the same feature for all surrounding elements, for everything in the auxiliary forms, and so on. I wrote about it in details in my book.

2. All the parameters of visibility must be easily controlled by users.

   Rules 1 and 2 are the projections of the full users' control over a program on the different sets of parameters. The first rule deals with the locations and sizes; the second rule deals with the colors, fonts, and some auxiliary things.

3. The users' commands on moving / resizing of objects or on changing the visibility parameters must be implemented exactly as they are; no additions or expanded interpretation by developer are allowed.

   It can be a bit strange at the beginning of new design to control yourself and not to add anything of your own to the users' commands. You may be a designer with many years of practice; you really know what must be done to make the view of an application better in this or that situation. But this is another world; the full users' control over an application must be really full; you do not leave anything for yourself as a second control circuit. Eventually you will find that nobody needs your super control from behind the curtains. Where you have to apply all your skills (the higher – the better) is in designing the default view. The highest credit to you is the big percentage of users, who would not change anything at all but work exactly with your proposed design. Yet, the possibility of all those moving, resizing, and tuning must be there for users to try them at the first wish.

4. All the parameters must be saved and restored.

   Saving the parameters for restoring them later is definitely not a new thing and is practiced for many years. But only the passing of the full control to users and the significant increase of the number of parameters that can be changed by users turned this saving / restoring of parameters from the feature of the friendly interface into a mandatory thing. If user has spent some time on rearranging the view to whatever he prefers, then the loss of these settings is inadmissible. As the full users' control means the possibility of changing any parameter, then the saving / restoring of all the parameters must be implemented.

5. The above mentioned rules must be implemented at all the levels beginning from the main form and down to the farthest corners.

The basis of the last rule is also an explanation of an automatic change in user's view whenever anyone starts to work with user-driven applications. The new applications are visually indistinguishable from the standard so the users, even informed about the new features, start to work with them as usual but quickly find the new features and their advantages. The movability of all the screen elements is so helpful and valuable in the majority of applications that users immediately begin to use it. They got so used to this total movability that expect it everywhere; they automatically try to move and resize everything not only in the main form of an application but in all the auxiliary forms; that is why the above mentioned rules must be applied to design on all the levels. This expectation of total movability is so strong that users are really disappointed when in parallel with the new applications they continue to work with the old well known programs and find out that those programs didn't get the new features in some magic way but still contain unmovable and non-resizable elements. The disappointment is really big.

## Conclusion

There was no conclusion in the preliminary version of this article, but then I got several remarks from reviewers and those remarks emphasized some often asked questions about my ideas. I decided that answering these remarks and questions would be the best conclusion to the article.

1. *Because this is all about the interface design then it is the prerogative of specialists on design to make a decision about the viability of these new ideas.*

Not interface designers but only users have to decide whether they want to use this movability of the screen elements or not. This decision has to be made not by popular voting but each user has to do it personally. Even if you are an excellent interface designer, look at the problem as a user, because you are definitely a user of many programs. Interior of your house, things on your desk, and the objects on the screen of your computer – these are the parts of your own world. Their configuration is not decided by city or country voting on election day. The first two things are not argued by anyone, though it wasn't always this way, but about the third thing people have doubts. The source of these doubts is obvious: it happens only because users were never given a chance to try but were always told that only specialists can make the decisions about the interface. Remember how many times you were mad with interface designed by very good specialists? Try the applications which you can rearrange yourself in any way you want (in addition to very good default views), and only then make your decision about the total movability of elements in the programs.

2. *Specialists on interface design know better than anyone else how a good interface must look like, so only such specialists must be allowed to make a decision about the applications' views.*

I strongly believe in specialization and always prefer specialists to do the job instead of amateurs. But the decision about the use of the work force must be made only voluntary and only by the person who needs some work to be done. (Life threatening jobs are the exceptions, but we are not discussing such issues.) If you organize a party for several dozens or more people then chances are high that you will hire some specialist to organize everything in order, but I doubt that you need to call a waitress from a nearby restaurant to serve a quick meal for your kids before they rush to school. The majority of people are qualified enough to put plates, spoons, forks, knives, and cups on the table for their ordinary meal and rarely wait for somebody's direct orders for such action.

The same with the applications. There are some very special programs in which the placement of the screen objects is so essential that it can be done only after long considerations and better not to be changed. But the majority of programs deal with objects that can be placed anywhere and resized in any way without doing any harm to the application itself. Such programs can be easily changed by users; in addition, there is always a good default view which can be reinstalled at any moment.

I never declare a rule "either designer or users". Each program has to have a good professional design, but users must get an opportunity to change programs in any way they want.

3. *The end users want to accomplish the task that the application is intended to support rather than configure the interface.*

Certainly, the main goal of using any program is to accomplish some task. But the variety of tasks is so wide that they cannot be solved with the fixed interface, so there are decades of work on the ways to adjust interface to the particular tasks. Users are also very different in their eagerness to spend time on interface changes. For those, who do not want to spend even an extra second, the movability is not needed at all; such users always work with the provided configuration and for this group of users the movability can be simply switched off. In user-driven applications the movability is always available on the first wish <u>of each user</u>. It is not a feature which is imposed on everyone as a burden; it is a new possibility that is provided in full but is used by each user individually at his wish. As the rules are few and primitive – everything is movable by any inner point; if resizable, it is done by any border point – then no new learning is needed except the knowledge of the fact that everything is movable.

4. *Experiments are conducted to test which alternative interfaces are best for the majority of users; programs are designed according to the results of these experiments.*

Suppose that interface is organized according to such experiments; what are you going to do in case you do not belong to this majority and has another opinion? A program is used not by average user but by each individual. Currently used programs are supposed to satisfy an average user; the new applications satisfy each one. I don't see how the first case can be better than the second. What do you prefer as a user of any program: the information that this program has the best view for majority or an application that is always organized according to your wish even if your demands can significantly change from one moment to another?

5. *The computer should observe user's work, anticipate the needs of the user, and take on work that the user has not even asked for yet.*

I absolutely disagree with such view; I don't want programs to do anything on their own. There are questions on which I am not sure; in such situations <u>I can ask</u> a program to do something of its own. In each of those cases there are some variants which were coded by developer and there is a predefined algorithm of selection. In all other cases user has to make a decision. In any situation computer (program) must do only what it was asked to do but nothing of its own.

6. *It is very annoying that applications often change their layout and then users have to go hunting for the data they need.*

    I write all the time that developers of programs must be banned from ruling the view. Developers must provide the information but only users (each one personally) have to decide about the view. Developers can propose the new views that they consider to be better than previous but only user has to decide whether to accept the new view or not. In this way there would be no searching for the needed data which developer has hidden somewhere else in the new version. All the objects would be exactly in the places which user ordered them to be.

    If you cut developers from controlling the view of applications, then the full control is passed to users. In order to change the view in an arbitrary way users need an easy way of moving / resizing all the elements. My work is just about the algorithm and its implementation. I hope you try the new applications and find for yourself how the movability of all the screen objects changes our work with computers.

## References

1. H.Lieberman, F.Paterno, V.Wulf, *End-User Development*. Springer, 2006
2. S.J.Maloney, R.Smith, Directness and Liveness in the Morphic User Interface Construction Environment, UIST'95, pp.21-28 (November 1995).
3. I.E.Sutherland, SketchPad, A Man-Machine Graphical Communication System, 1963  Can be found at http://www.cl.cam.ac.uk/techreports/UCAM-CL-TR-574.pdf
4. S.Andreyev, *World of Movable Objects*, MoveableGraphics project, SourceForge, 2010; http://sourceforge.net/projects/movegraph/files
5. S.Andreyev, *Into the World of Movable Objects.* Computing in Science and Engineering, v.13, Issue 4, pp.79 - 84, July 2011